\documentclass[11pt]{article}

\usepackage{jheppub}

\usepackage{amsmath}
\usepackage{amssymb}
\usepackage{amsfonts}
\usepackage{mathrsfs}

\usepackage{mathtools}

\usepackage{tikz}
\usetikzlibrary{decorations.pathmorphing,backgrounds}
\tikzset{zigzag/.style={decorate, decoration=zigzag}}
\makeatletter
\def\@hex@@Hex#1%
 {\if a#1A\else \if b#1B\else \if c#1C\else \if d#1D\else
  \if e#1E\else \if f#1F\else #1\fi\fi\fi\fi\fi\fi \@hex@Hex}
\makeatother

\usepackage{xcolor}

\definecolor{plum}{rgb}{0.36078, 0.20784, 0.4}
\definecolor{chameleon}{rgb}{0.30588, 0.60392, 0.023529}
\definecolor{cornflower}{rgb}{0.12549, 0.29020, 0.52941}
\definecolor{scarlet}{rgb}{0.8, 0, 0}
\definecolor{brick}{rgb}{0.64314, 0, 0}	

\newcommand{\ts}[1]{\textrm{\tiny #1}}
\newcommand{\ms}[1]{\textrm{\tiny $#1$}}
\newcommand{\eps}{\epsilon}

\newcommand{\bX}{\bar{X}}

\newcommand\nts{\negthickspace}
\newcommand\bns{\nts \nts \nts}

\DeclareMathOperator{\extdm}{d}
\newcommand{\extd}{\extdm \!}

\newcommand{\eq}[2]{\begin{equation}
#1\label{#2}
\end{equation}}

\preprint{TUW--20--04}

\newcommand{\mytitle}{Universal flow equations and chaos bound saturation in 2d dilaton gravity}

\title{\boldmath \mytitle}

\author[a]{D.~Grumiller}
\author[b]{and R.~McNees}

\affiliation[a]{Institute for Theoretical Physics, TU Wien, Wiedner Hauptstrasse 8--10, A-1040 Vienna, Austria}
\affiliation[b]{Department of Physics, Loyola University Chicago, Chicago, IL, USA}

\emailAdd{grumil@hep.itp.tuwien.ac.at}
\emailAdd{rmcnees@luc.edu}

\abstract{We show that several features of the Jackiw--Teitelboim model are in fact universal properties of two-dimensional Maxwell-dilaton gravity theories with a broad class of asymptotics. These theories satisfy a flow equation with the structure of a dimensionally reduced $T$\,$\bar{T}$ deformation, and exhibit chaotic behavior signaled by a maximal Lyapunov exponent. One consequence of our results is a no-go theorem for smooth flows from an asymptotically AdS$_2$ region to a de Sitter fixed point.
}

\begin{document}
\maketitle

\section{Introduction}
\label{sec:Intro}

Chaotic systems are characterized by their Lyapunov exponents $\lambda_L$, which give a characteristic inverse time-scale and determine the rate at which nearby trajectories separated by $\delta z(t)$ in phase space diverge exponentially from each other, 
\eq{
\delta z(t)\sim e^{\lambda_L t} \, \delta z(0)\,. 
}{eq:chaos0}
At finite temperature $T$ there is a conjectured bound on chaos \cite{Maldacena:2015waa},
\eq{
\lambda_L \leq 2\pi\, T
}{eq:chaos1}
which is saturated by (large-$N$) field theories dual to Einstein gravity with negative cosmological constant. Saturation of the chaos bound \eqref{eq:chaos1} for a CFT is thus a signal that it may have an AdS dual.

What can be said beyond AdS/CFT? It is of general interest to find out which features observed so far are specific to AdS/CFT and which are universal aspects of holography. The universality of the chaos bound \eqref{eq:chaos1} as well as its saturation for holographic systems could be a helpful piece in this puzzle.

In order to address this question efficiently we need a class of gravity models that is simultaneously rich enough to allow for a variety of different asymptotics and simple enough to allow quantitative analysis. Two-dimensional (2d) dilaton gravity seems tailor-made for this task, for two reasons. First, there is a concrete holographic proposal relating the quantum mechanical Sachdev--Ye--Kitaev (SYK) model \cite{Kitaev:15ur,Sachdev:1992fk,Sachdev:2010um} with the gravitational Jackiw--Teitelboim (JT) model \cite{Teitelboim:1983ux,Jackiw:1985je} where chaos bound saturation was established \cite{Maldacena:2016hyu,Jensen:2016pah}. Second, two-dimensional dilaton gravity allows for a variety of models with different asymptotics that can be treated on equal footing, including asymptotically AdS$_2$, flat space, and other behavior, see e.g.~\cite{Grumiller:2002nm}. 

We stress that the following universality argument is too naive: all 2d geometries are locally conformally flat and hence locally conformally AdS$_2$; thus, one should expect universal features for all 2d gravity theories, their asymptotic symmetries, their boundary theories and the holographic Lyapunov exponents. A counter example to this argument is SYK/JT vs.~the flat space holographic model studied in \cite{Afshar:2019axx}: the respective asymptotic symmetries differ and the boundary theories, described by Schwarzian vs.~twisted warped action, differ as well. Thus, while it still may be true that the holographic Lyapunov exponents turn out to be universal, this is not a result that should be anticipated based on the naive argument above.

At a technical level our goal is to generalize the discussion in the first half of \cite{Gross:2019ach} to generic 2d Maxwell-dilaton gravity. Thus, we continue with a brief summary of their work.

The starting point of \cite{Gross:2019ach} was quantum mechanics, which is non-universal in the IR, since every (regular, non-derivative) operator is relevant, but rather universal in the UV as a consequence of the paucity of irrelevant operators. The IR landscape can be investigated by integrable deformations of quantum mechanics that deform the action $\Gamma$ through the flow equation
\eq{
\frac{\partial\Gamma}{\partial\lambda} = F(t,j;\,\lambda)
}{eq:chaos2}
where $\lambda$ is the deformation-parameter (or flow-parameter), $t$ is the stress-scalar (i.e., the 1-dimensional version of the stress tensor), $j$ denotes other physical quantities (e.g.~a $u(1)$ charge) and $F$ is some functional thereof. The particular choice of $F$ made in \cite{Gross:2019ach} was in line with $T\bar T$-deformations of CFT$_2$ \cite{Zamolodchikov:2004ce,Smirnov:2016lqw,Cavaglia:2016oda}. More precisely, this irrelevant deformation generates on the gravity side a radial cutoff on the asymptotically AdS$_2$ geometry, which can be understood by dimensionally reducing from three to two dimensions on the gravity side and from CFT$_2$ to quantum mechanics on the field theory side. The explicit form of the flow equation \eqref{eq:chaos2} is
\eq{
\frac{\partial \Gamma_c}{\partial \lambda} =  2\,\int \extd\tau \sqrt{\gamma}\,\frac{E_{\textrm{\tiny{FT}}}^2 - J_{\textrm{\tiny{FT}}}^2}{1-4\,\lambda\,E_{\textrm{\tiny{FT}}}} = 2\,\int \extd\tau \sqrt{\gamma}\,\frac{(T^{\tau}{}_{\tau})^2 + T^{\tau\phi}\,T_{\tau\phi}}{1-4\,\lambda\,T^{\tau}{}_{\tau}}
}{eq:chaos3}
where $\Gamma_c$ is the on-shell action with finite cutoff surface, $\lambda$ is the deformation parameter, $\tau$ is the coordinate along the boundary with induced volume form $\sqrt{\gamma}$, $E_{\textrm{\tiny{FT}}}$ is the field theory energy and $J_{\textrm{\tiny{FT}}}$ is essentially the field theory $u(1)$ charge. By dimensional oxidation one can interpret these quantities as components of a CFT$_2$ stress tensor, $T^\tau{}_\tau=E_{\textrm{\tiny{FT}}}$ and $T^{\tau\phi}=T_{\tau\phi}=iJ_{\textrm{\tiny{FT}}}$, which shows that the deformations induced by \eqref{eq:chaos3} are a dimensionally reduced variant of $T\bar T$-deformations. The effects of chaos were studied for these deformed theories on the gravity side by adapting the discussion of \cite{Engelsoy:2016xyb} to finite cutoff. The essence of that calculation is captured by Fig.~\ref{fig:1} in the body of our paper and will be reviewed briefly there. Conceptually, one calculates a time delay of an outgoing signal towards the boundary that is caused by a matter shock wave released from the boundary. Technically, this boils down to a simple evaluation of lengths of certain geodesics. The outcome is a time delay formula reminiscent of \eqref{eq:chaos0} with a Lyapunov exponent satisfying the chaos bound \eqref{eq:chaos1}. This shows that the result for the Lypunov exponent applies not only to JT but also deformations thereof.

Our main goal is to generalize the discussion summarized in the paragraph above to generic 2d Maxwell-dilaton gravity theories, including models that do not asymptote to AdS$_2$. In particular, we intend to establish some flow equations and calculate on the gravity side the Lyapunov exponent to check whether or not it obeys or saturates the chaos bound  \eqref{eq:chaos1}.

This paper is organized as follows. In section \ref{sec:Review} we review salient features of generic (Euclidean) 2d Maxwell-dilaton gravity in the presence of an asymptotic boundary or a finite radial cutoff. In section \ref{sec:TbarT} we apply $T\bar T$-like deformations to all these models and derive the flow equations analogous to \eqref{eq:chaos3}. In section \ref{sec:dS} we provide a no-go result that forbids smooth flows from AdS$_2$ to an IR dS$_2$ fixed point, and consider possible loopholes in our argument. In section \ref{sec:Chaos} we address chaos and calculate on the gravity side the Lyapunov exponent, establishing saturation of the chaos bound \eqref{eq:chaos1}. In section \ref{sec:Discussion} we conclude.

\section{Mini-review of Maxwell-dilaton gravity in two dimensions}
\label{sec:Review}

In this section we give a brief review of relevant aspects of (Euclidean) 2d Maxwell-dilaton gravity. In section \ref{sec:2.1} we display the bulk action and generic solutions (in a suitable gauge), and summarize some of their properties. In section \ref{sec:2.2} we consider boundary issues and present the full action. In section \ref{sec:2.3} we use the Euclidean action to extract the free energy and other thermodynamical quantities of interest.

\subsection{Bulk action and generic solutions}\label{sec:2.1}

Maxwell-dilaton gravity in two dimensions with Euclidean signature has a bulk action 
\eq{
  I[g_{\mu\nu},\,A_\mu,\,X] = - \frac{1}{2\,\kappa^{2}}\,\int_{M} \nts\nts \extd^{2}x\sqrt{g}\,\Big(X\,R -U(X)\,(\nabla X)^2 - 2\,V(X) - 2\,f(X)\,F^{\mu\nu} F_{\mu\nu}\Big)
}{BulkAction}
where $\kappa$ is the gravitational coupling constant, $X$ is the dilaton, $g_{\mu\nu}$ is the metric on $M$, and $F_{\mu\nu}$ is the field strength for an Abelian gauge connection $A_{\mu}$. The kinetic potential $U$, dilaton potential $V$ and dilaton-Maxwell coupling $f$ are arbitrary functions of the dilaton at this stage.

Generic solutions of these theories, expressed here in Schwarzschild-like coordinates and axial gauge 
\eq{
    \extd s^{2} = \xi\,\extd\tau^{2} + \frac{1}{\xi}\,\extd r^{2} \qquad\qquad A_{\mu}\,\extd x^{\mu} = A_{\tau}\, \extd\tau 
}{DGSolution} 
obey a generalized Birkhoff theorem, i.e., they have a Killing vector $\partial_{\tau}$ whose orbits are isosurfaces of $X$. This means $\xi$ and $A_\tau$ can be expressed as functions of $X$, satisfying
\begin{align} 
    \xi(X) &= e^{Q(X)}\,\left(w(X) - 2\,M + \frac{1}{4}\,q^{2}\,H(X)\right)
    \label{MetricSolution} \\ \label{VectorSolution}
    A_{\tau}(X) &= -\frac{q}{4}\,\Big(H(X) - H(X_h)\Big) + A_{\tau}(X_h) 
    \\ \label{DilatonSolution}
     \partial_{r} X &= e^{-Q(X)} 
\end{align}
where $Q$, $w$, and $H$ are integrals of the functions appearing in \eqref{BulkAction}
\begin{align}
    Q(X) &= Q_0 + \int^{X} \bns \extd\tilde{X} \,U(\tilde{X}) \label{Qdef} \\ \label{omegadef}
    w(X) &= w_0 - 2\int^{X} \bns \extd\tilde{X}\,V(\tilde{X})\,e^{Q(\tilde{X})} \\ \label{Hdef}
    H(X) &= H_0 + \int^{X} \bns \extd\tilde{X}\,\frac{e^{Q(\tilde{X})}}{f(\tilde{X})} ~.
\end{align}
Without loss of generality we set the additive constant $w_0$ to zero by absorbing it into the constants of motion $M$ and $q$,
which are interpreted as mass and charge parameter, respectively. The constant $Q_0$ is set to zero by a rescaling of the coordinates. The constant $H_0$ is irrelevant because it appears alongside $w_0$ in \eqref{MetricSolution} and drops out of the difference $H(X)-H(X_h)$ in the abelian connection \eqref{VectorSolution}, and hence also is set to zero. The only curvature invariant, the Ricci scalar, depends on the three functions \eqref{Qdef}-\eqref{Hdef} and their derivatives,
\eq{ 
R = -\frac{\partial^2\xi}{\partial r^2} = -e^{Q(X)}\,\partial_X\Big(w'(X)+\frac14\,q^2\,H'(X) + Q'(X)\,\big(w(X)-2M+\frac14\,q^2\,H(X)\big)\Big)\,.
}{eq:chaos8}

For some models there is an isolated sector of non-generic solutions, so-called constant dilaton vacua, where the dilaton $X=X_\text{cdv}=\rm const.$ is a solution of the non-differential equation $V(X_\text{cdv})=q^2/(8f(X_\text{cdv}))$ and the Ricci scalar is a constant given by
\eq{
\textrm{constant\:dilaton\;vacua:}\qquad R = 2V'(X_\text{cdv}) + \frac{q^2f'(X_\text{cdv})}{4f^2(X_\text{cdv})}\,.
}{eq:chaos6}
These special solutions are not generic geometries, and in the present work they will only be relevant when discussing properties of IR fixed points of the flows discussed in section \ref{sec:TbarT}.

Instead, we consider generic solutions of models where the metric is non-negative on the semi-infinite interval of positive dilaton values $X_h \leq X < \infty$, with $X_h$ given by the Euclidean horizon condition $\xi(X_h) = 0$. If $\xi(X)=0$ has multiple roots then $X_h$ is taken to be the largest such solution, and if there are no such roots then the interval is $X\in(0,\,\infty)$.\footnote{In section \ref{sec:4.3} we will consider an extension of these models that allow the lower end of the interval to extend to negative values.} The function $e^{Q(X)}$ is assumed to be non-zero on this interval, so that the location of $X_h$ is given by the condition
\eq{
    w(X_h) + \frac{1}{4}\,q^2\,H(X_h) = 2\,M ~.
}{eq:1}
Regularity of the solutions at $X_h$ requires that the Euclidean time has periodicity $\tau \sim \tau + \beta$ given by
\eq{
    \beta = \frac{4\pi}{\partial_r \xi}\Big|_{r_h}
    = \frac{4\pi}{w'(X_h)+\tfrac{1}{4}\,q^2\,H'(X_h)} ~.
}{eq:2}
For the models we are interested in the function $w(X)$ tends to $+\infty$ in the limit $X \to \infty$, and we assume that this behavior, rather than that of $H(X)$, determines the asymptotics of $\xi(X)$. Thus any dilaton-Maxwell coupling $f(X)$ is allowed as long as $H(X)$ satisfies $H(X)/w(X) \to 0$ in the limit $X \to \infty$. Under these conditions the $X \to \infty$ behavior of $\xi(X)$ is dominated by the first term in \eqref{MetricSolution}, which we denote by
\eq{
    \xi_{0}(X) = e^{Q(X)}\,w(X) ~.
}{eq:chaos7}
The function $\xi_{0}(X)$ is equivalent to $\xi(X)$ with both mass $M$ and charge $q$ set to zero, so the physical state corresponding to this solution is sometimes referred to as ``ground state''. 

\subsection{Boundary issues and full action}\label{sec:2.2}

For theories on a manifold with boundary, a variational principle with Dirichlet conditions on the fields requires an additional term in the action. With our parameterization of the dilaton this Gibbons--Hawking--York boundary term \cite{Lau:1996fr} takes the form
\begin{gather}\label{GHYTerm}
    I_{\textrm{\tiny{GHY}}} = - \frac{1}{\kappa^{2}}\,\int_{\Sigma} \nts \extd^{1}x\sqrt{h}\,X\,K ~.
\end{gather}
But for theories defined on a spacetime with spatial infinity (which in the Lorentzian continuation could correspond to null or spatial infinity), the construction of the variational principle is more subtle. In that case we introduce a boundary by cutting off the spacetime at some large value $X_c$ of the dilaton, and supplement \eqref{BulkAction} and \eqref{GHYTerm} with an additional ``boundary counterterm'' on the cut-off surface $\Sigma$. The appropriate boundary term was constructed in \cite{Grumiller:2007ju} for theories with the asymptotics described above and yields the full action
\begin{align} \label{FullAction}
    \Gamma = &\,\, - \frac{1}{2\,\kappa^{2}}\,\int_{M} \nts\nts \extd^{2}x\,\sqrt{g}\,\Big(X\,R -U(X)\,(\nabla X)^2 - 2\,V(X) - 2\,f(X)\,F^{\mu\nu} F_{\mu\nu}\Big) \\ \nonumber
    & \,\, - \frac{1}{\kappa^{2}}\,\int_{\Sigma} \nts \extd^{1}x\,\sqrt{h}\left( X\,K - e^{-Q(X)}\,\sqrt{\xi_0} \,\right) ~.
\end{align}
Henceforth we will set $\kappa^2 = 8\pi G_2 = 1$. The action for the full spacetime is recovered in the $X_c \to \infty$ limit of \eqref{FullAction}. This construction gives a well-defined variational formulation of the theory, in the sense that the variation $\delta \Gamma$ vanishes on-shell for all field variations with the same asymptotic fall-off as the sub-leading terms in the solutions \eqref{MetricSolution}-\eqref{DilatonSolution}.

The main limitation of the action \eqref{FullAction} is that we assumed the boundary to be a dilaton iso-surface. In the present work we focus on applications where the action \eqref{FullAction} can be used. We address generalizations in the concluding section \ref{sec:Discussion}.

\subsection{Thermodynamics}\label{sec:2.3}

The Euclidean action \eqref{FullAction} evaluated on solutions of the form  \eqref{MetricSolution}-\eqref{DilatonSolution} describes the thermodynamics of Maxwell-dilaton gravity in an ensemble that, depending on the boundary conditions, may contain black holes, ``hot empty space'' solutions, and more exotic field configurations. The variational principle places Dirichlet conditions on the metric, gauge field, and dilaton at spatial infinity, which corresponds to a thermodynamic ensemble that fixes the proper local temperature, proper electrostatic potential, and a conserved dilaton charge. In general, the thermodynamic ensemble only exists for some values of the cut-off $X_c$ that defines the surface $\Sigma$ in \eqref{FullAction}. The surface is regarded as a cavity wall where the system is coupled to a thermal reservoir which maintains the proper temperature and electrostatic potential, given by 
\begin{align}
    T_c &= \beta_{c}{}^{-1} = \frac{1}{\sqrt{\xi_c}}\,\beta^{-1}\\
    \Phi_c &= \frac{A_{\tau}(X_c) - A_{\tau}(X_h)}{\sqrt{\xi_c}} ~.
\end{align}
The dilaton charge is given by $D_c=D(X_c)$, where any well-behaved function of the dilaton $D(X)$ gives such a conserved charge \cite{Gibbons:1992rh}. A simple choice is $D(X) = X$, but below we shall make the more convenient choice $D(X)=1/[4w(X)]$ and hence
\eq{
D_c = \frac{1}{4w(X_c)}\,.
}{eq:Dc}
The thermodynamic potential for the ensemble is obtained from the on-shell action $\Gamma_c$ by
\begin{gather}
    \Gamma_{c} = \beta_c\,Y_{c}(T_c, \Phi_c, D_c)
\end{gather}
with $Y_c$ related to the Helmholtz free energy $F_c$ and the internal Energy $E_c$ by appropriate Legendre transforms
\begin{gather}
    Y_{c}(T_c, \Phi_c, D_c) = F_{c}(T_c, D_c, q) - q\,\Phi_c = E_{c}(S, D_c, q) - T_c\,S - q\,\Phi_c ~.
\end{gather}
There is always some range of values of the cut-off $X_c\in(X_h,\,X_c^{\textrm{\tiny{max}}})$ for which the specific heat is positive and the ensemble is well-defined. In some cases $X_c^{\textrm{\tiny{max}}}=\infty$ so that the ensemble remains well-defined as the system is decoupled from the reservoir. Examples for such behavior are the exact string black hole \cite{Dijkgraaf:1992ba} and sufficiently large spherically symmetric AdS black holes in Einstein gravity in dimension $D\geq 3$. But for many theories and boundary conditions $X_c^{\textrm{\tiny{max}}}$ is finite so that the specific heat becomes negative when $X_c>X_c^{\textrm{\tiny{max}}}$ and the canonical ensemble is not defined. Examples for such behavior are so-called Minkowski ground state models, including Schwarzschild--Tangherlini black holes in $D$-dimensional spherically symmetric Einstein gravity. %
See \cite{Grumiller:2007ju} for a discussion of the thermodynamics of these and other models.

Evaluating the action for solutions \eqref{MetricSolution}-\eqref{DilatonSolution} yields
\begin{gather}\label{OnShellAction1}
    \Gamma_c = \beta_{c}\, \Big( e^{-Q_c} \big(\sqrt{\xi_0} - \sqrt{\xi_c} \big) - 2\pi X_{h} T_{c} - q\,\Phi_c \Big)
\end{gather}
where $\xi_0$, defined in \eqref{eq:chaos7}, is also evaluated at the cut-off. The first law yields an entropy $S = 2\pi X_h$, a conserved electric charge given by the parameter $q$ appearing in the gauge field \eqref{VectorSolution}, and internal energy
\begin{gather}\label{BulkInternalEnergy}
    E_c = e^{-Q_{c}} \big(\sqrt{\xi_0} - \sqrt{\xi_c} \big) ~.
\end{gather}
This last result agrees with the conserved quantity obtained via both Hamiltonian methods and the Brown--York quasi-local stress tensor.

\section{\texorpdfstring{$\boldsymbol{T\,\bar T}$}{T T} flow equations}
\label{sec:TbarT} 

In this section we consider flows \eqref{eq:chaos2} of the action \eqref{FullAction} as we move the cutoff $X_c$ from larger to smaller values. The holographic interpretation (if available) is that in the dual quantum mechanical theory we flow from the UV to the IR. The main result of this section is that Maxwell-dilaton gravity theories considered in the previous section naturally satisfy a flow equation like \eqref{eq:chaos3}. This flow can be understood as a consequence of the quasi-local form of the first law of black hole thermodynamics, with certain thermodynamic variables held fixed.

In section \ref{sec:3.1} we introduce field theory quantities, some of which are held fixed in the flows considered in section \ref{sec:3.2}, where we derive our main result, the flow equation. In section \ref{sec:3.3} we discuss some properties of and provide an interpretation for the flow equation.

\subsection{Field theory quantities}\label{sec:3.1}

To show the main result we switch over to the ``field theory variables'' employed in \cite{Hartman:2018tkw}. The field theory geometry is described by a one-dimensional metric $\gamma$ that is related to the metric on the cut-off surface by scaling out the factor $\xi_0$ that characterizes the asymptotic behavior
\begin{gather}
    \xi_c = \xi_0\,\gamma ~.
\end{gather}
The proper thermodynamic quantities $E_c$, $T_c$, and $\Phi_c$ measured by a bulk observer at $X_c$ are then related to the associated field theory quantities by appropriate factors of $\sqrt{\xi_0}$.
\eq{
    E_\ts{FT} = \sqrt{\xi_0}\,E_c \qquad\quad
    T_\ts{FT} = \sqrt{\xi_0}\,T_c \qquad\quad
    \Phi_\ts{FT} = \sqrt{\xi_0}\,\Phi_c \qquad\quad 
    \beta_\ts{FT} = \frac{1}{\sqrt{\xi_0}}\,\beta_c 
}{FieldTheoryQuantities}
Field theory and cut-off quantities coincide only for ``Minkowski ground state models,'' $\xi_0=1$. In all other cases \eqref{FieldTheoryQuantities} relates them by cut-off dependent (but state-independent) factors. 

The on-shell action \eqref{OnShellAction1} expressed in terms of field theory quantities,
\begin{gather}\label{ActionFTQuantities}
    \Gamma_c = \beta_\ts{FT}\,Y_\ts{FT} = \beta_\ts{FT}\,\big(\,E_\ts{FT} - S\,T_\ts{FT} - q\,\Phi_\ts{FT} \big) 
\end{gather}
leads to the same thermodynamic interpretation as before, with the entropy $S = 2\pi X_h$ and conserved electric charge $q$ unchanged. In particular, the field theory thermodynamic potential $Y_\ts{FT}$ and energy $E_\ts{FT}$ satisfy the appropriate forms of the first law
\begin{gather}\label{eq:1stlaw}
    \extd Y_\ts{FT} = - S \extd T_\ts{FT} - q\, \extd \Phi_\ts{FT} - \psi_\ts{FT}\,\extd D \\
    \extd E_\ts{FT} = T_\ts{FT}\,\extd S + \Phi_\ts{FT}\,\extd q - \psi_\ts{FT}\,\extd D 
\end{gather}
for any choice of dilaton charge $D$.

\subsection{Dilaton cut-off flows}\label{sec:3.2}

We consider now flows that hold the field theory geometry and electric field fixed, which in particular means fixing the temperature $T_\ts{FT}=1/\beta_\ts{FT}$ and the electrostatic potential $\Phi_\ts{FT}$. The key quantity that changes along such flows is the cut-off value of the dilaton, $X_c$. Since the theories we consider have $w(X) \to \infty$ as $X \to \infty$, there is always some open interval $[X_\text{min},\infty)$ on which $w(X)$ is monotonically increasing. In some models the lower end of this interval may extend to the horizon; otherwise it will occur at a local minimum of $w(X)$. For $X$ in this interval we define the ``universal cut-off'' $\lambda$ via
\eq{
    w(X_c) = \frac{1}{4\,\lambda}
}{eq:chaos5}
and replace dependence on $X_c$ with dependence on $\lambda$. Then the field theory metric can be written as
\eq{
    \gamma = 1 - 8\,M\,\lambda + q^{2}\lambda\,H(\lambda) 
}{eq:chaos9}
and the relation \eqref{FieldTheoryQuantities} between $E_c$ and $E_\ts{FT}$ yields
\begin{gather}\label{FTEnergy}
    E_\ts{FT} = \frac{1 - \sqrt{\gamma}}{4\,\lambda} = \frac{1 - \sqrt{1 - 8\,M\,\lambda + q^{2}\lambda\,H(\lambda)}}{4\,\lambda} ~.
\end{gather}
In \cite{Gross:2019ach} this was viewed as a consequence of the flow equation; here it is the result of the bulk geometry and black hole thermodynamics.\footnote{%
Indeed, this notion of energy was used in Eq.~(3.24) of \cite{Grumiller:2007ju} when considering how internal energy and gravitational binding energy are related to the ADM mass.} 

The flow equation is obtained by considering the response of the on-shell action to a small change in the universal cut-off $\lambda$ with $\beta_\ts{FT} = 1 / T_\ts{FT}$ and $\Phi_\ts{FT}$ held fixed. Making the choice \eqref{eq:Dc}, $D = \lambda$, for the dilaton charge, the first law \eqref{eq:1stlaw} implies
\eq{
    \frac{\partial \Gamma_c}{\partial \lambda}\Big|_{\beta_\ts{FT},\Phi_\ts{FT}} = \beta_\ts{FT}\,\frac{\partial E_\ts{FT}}{\partial \lambda}\Big|_{S,q} = \int \extd\tau \sqrt{\gamma }\,\frac{\partial E_\ts{FT}}{\partial \lambda}\Big|_{S,q} ~.
}{eq:angelinajolie}
In the last step we have expressed $\beta_\ts{FT}$ as the integral over $\tau$ with an appropriate factor of $\sqrt{\gamma}$. The derivative of $E_\ts{FT}$ is taken with the charge $q$ and entropy $S$ held fixed. Since $S = 2\pi X_h$ and $X_h$ is determined by \eqref{eq:1}, this is equivalent to holding $M$ and $q$ fixed when taking the derivative. 
From \eqref{FTEnergy} we then have
\begin{gather}\label{FirstFlow}
    \frac{\partial E_\ts{FT}}{\partial \lambda}\Big|_{S,q} = \frac{1}{1 - 4\lambda\,E_\ts{FT}}\,\left( 2\,E_\ts{FT}^{\,2} - \frac{1}{8}\,q^{2}\,\frac{\partial H(\lambda)}{\partial \lambda}\right) ~.
\end{gather}
The right-hand side of this equation is (minus) the dilaton chemical potential $\psi_\ts{FT}$ for the dilaton charge $D = \lambda$. The last term in parentheses can be re-written using the definitions \eqref{omegadef}-\eqref{Hdef} of $w$ and $H$ as
\begin{gather}
    \frac{\partial H(\lambda)}{\partial \lambda} = \frac{1}{8\,\lambda^{2}\,V(\lambda)\,f(\lambda)} ~.
\end{gather}
We can then express the derivative of the on-shell action along this flow as
\eq{
        \frac{\partial \Gamma_c}{\partial \lambda}\Big|_{\beta_\ts{FT},\Phi_\ts{FT}} = \int \extd\tau \sqrt{\gamma}\,\frac{2\,(E_\ts{FT}^{\,2} - J_\ts{FT}^{\,2})}{1-4\,\lambda\,E_\ts{FT}} ~,
}{UniversalFlowEquation}
where $J_\ts{FT}$ is defined as
\eq{
    J_\ts{FT} =  \frac{q}{\sqrt{128\,\lambda^{2}\,V(\lambda)\,f(\lambda)}} ~.
}{eq:chaos10}
For the dimensional reduction of the BTZ black hole to a two-dimensional Maxwell-dilaton gravity, the quantity $J_\ts{FT}$ is the constant angular momentum. This suggests the following identification with the components of the stress tensor in a three-dimensional theory
\begin{gather}\label{eq:Identifications}
    T^{\tau}{}_{\tau} = E_\ts{FT} \qquad\qquad T^{\tau\phi} = T_{\tau\phi} = i\,J_\ts{FT} ~.
\end{gather}
The flow equation then takes the form
\eq{
        \frac{\partial \Gamma_c}{\partial \lambda}\Big|_{\beta_\ts{FT},\Phi_\ts{FT}} = 2\,\int \extd\tau \sqrt{\gamma}\,\frac{(T^{\tau}{}_{\tau})^2 + T^{\tau\phi}\,T_{\tau\phi}}{1-4\,\lambda\,T^{\tau}{}_{\tau}}
}{FinalFlowEquation}
which coincides precisely with \eqref{eq:chaos3}. Thus, the sort of flow equation satisfied by the JT model seems to be a universal feature of all the 2d Maxwell-dilaton gravity theories considered in the previous section.

\subsection{Properties and interpretation of the flow equation}\label{sec:3.3}

The dimensional reduction of the BTZ sector of AdS$_3$ gravity is included in the theories we consider, as are theories that admit asymptotically AdS$_2$ spacetimes \cite{Witten:2020ert}. In the latter case, the quantity $J_\ts{FT}$ flows from its value at $\lambda = 0$ ($X_c \to \infty$) as the cut-off is adjusted. But in the course of deriving \eqref{FinalFlowEquation} we have not \textit{assumed} the existence of a holographic dual for the bulk theory. Yet the structure of the flow equation is precisely what one would expect for the $3 \to 2$ dimensional reduction of a gravitational theory dual to a $T \bar{T}$ deformed CFT. This is perhaps indicative of a more general class of theories with holographic duals that are not asymptotically AdS$_2$.

From the point of view of black hole thermodynamics, the flow has a straightforward interpretation. The system is coupled to a thermal reservoir at $X_c$, and the cavity wall is moved in while holding the field theory quantities $T_\ts{FT}$ and $\Phi_\ts{FT}$ fixed. Thus 
\begin{gather}
    \extd \Gamma_c\Big|_{T_\ts{FT},\Phi_\ts{FT}} = - \beta_\ts{FT}\,\psi_\ts{FT}\,\extd \lambda
\end{gather}
with the dilaton chemical potential playing a role similar to pressure. Since any Maxwell-dilaton gravity theory subject to our assumptions satisfies a first law 
\eqref{eq:1stlaw}, the flow equation itself does not provide any further constraints on the functions $U$, $V$, and $H$ that characterize the theory.

As explained in the introduction, local Weyl-equivalence is not sufficient to render all 2d gravity theories equivalent. However, it is enough to imply that theories with distinct asymptotics obey the same flow equation. Under a local Weyl-rescaling of the metric, $e^{Q}$ has conformal weight 2, $w$ and $H$ are invariant, and the proper internal energy $E_{c}$ transforms with conformal weight $-1$. Thus $E_\ts{FT} = \sqrt{\xi_0}\,E_c$, the metric $\gamma = \xi_c / \xi_0$, and the cut-off $\lambda = 1/(4w)$ are all invariant, and the flow equation \eqref{FirstFlow} does not change under a local Weyl-rescaling of the fields. 

An interesting consequence of the flow equation is a rather universal factor 2 between the field theory energy in the UV and in the IR, which we now derive. Assume that we have a model where $w$ is strictly monotonic and the cutoff is sent to the asymptotic boundary, $\lambda\to 0$, (this is what we refer to as `UV'; this assumption includes the JT model and deformations thereof).\footnote{
Not all models have black hole solutions with positive specific heat in the limit $\lambda\to 0$. 
} Taking the limit $\lambda\to 0$ in the field theory energy \eqref{FTEnergy}  yields
\eq{
E_\ts{FT}^{\textrm{\tiny{UV}}} = \lim_{\lambda\to 0} E_\ts{FT} = M\,.
}{eq:lalapetz}
This result is to be expected, since $M$ is the (appropriately normalized) mass characterizing our black holes solutions. We turn now to the IR. Assuming there is a horizon, the IR limit is defined as the cutoff surface approaching the horizon. The locus of the horizon is at $w(X_h)=2M-\tfrac14\,q^2 H(X_h)$, implying $\gamma\to 0$ in that limit. Taking the limit $\gamma\to 0$ in the field theory energy \eqref{FTEnergy} yields
\eq{
E_\ts{FT}^{\textrm{\tiny{IR}}} = \lim_{\gamma\to 0} E_\ts{FT} = 2M - \frac14\,q^2 H(X_h)\,.
}{eq:lalapetz2}
Thus, in the limit of vanishing charge we have the universal ratio
\eq{
\frac{E_\ts{FT}^{\textrm{\tiny{IR}}}}{E_\ts{FT}^{\textrm{\tiny{UV}}}}\bigg|_{q=0} = 2
}{eq:lalapetz3}
for all 2d dilaton gravity models (subject to the assumptions mentioned).

As shown in section \ref{sec:Chaos}, there is another behavior associated with asymptotically AdS$_2$ black holes that generalizes to models with a broader class of boundary conditions. But first, we consider an interesting application of the flow equation derived above.

\section{A no-go result for flows to de Sitter fixed points}
\label{sec:dS}

In this section we consider an intriguing setup, namely a situation where we flow from AdS$_2$ in the UV (of the field theory, i.e., for large radii on the gravity side) to a dS$_2$ IR fixed point (again of the field theory, so for smaller radii on the gravity side) by virtue of the flow equations derived in the previous section. If such a flow existed it would be a smooth realization of the setup considered in \cite{Anninos:2017hhn}. We show now that no such flow exists.

To reduce clutter we set $f(X)=0$ in this section, but note that our results generalize trivially to $f(X)\neq 0$ by replacing $V\to V - q^2/(8f)$.

In section \ref{sec:4.1} we derive the no-go result. In section \ref{sec:4.2} we consider a particular example and generic aspects of the no-go result. 
In section \ref{sec:4.3} we show that, while a true dS$_2$ fixed point is inaccessible, a smooth flow might still approach a `quasi-fixed point' where the curvature is positive.

\subsection{Derivation of no-go result}\label{sec:4.1}

The main issue is the behavior of the function $w(X)$ defined in \eqref{omegadef}. Asymptoting to AdS$_2$ means that for large values of the dilaton this function has to be quadratic and positive, $w(X)=X^2/\ell^2 + \dots$. In order to have a fixed point in the IR the flow equation needs to stop, which happens once an extremum of $w(X)$ is reached. 

To show this, reconsider the flow equation as a differential equation with respect to the radial cutoff $r_c$,
\eq{
\frac{\partial\Gamma_c}{\partial r_c} = \frac{\partial\Gamma_c}{\partial w_c}\,\frac{\partial w_c}{\partial X_c}\,\frac{\partial X_c}{\partial r_c} = -2\,\frac{\partial\Gamma_c}{\partial w_c}\,V(X_c)
}{eq:chaos4}
where we used the defining relation \eqref{omegadef} and the solution for the dilaton \eqref{DilatonSolution}. An IR fixed-point, $\frac{\partial\Gamma_c}{\partial r_c} = 0$, arises for finite or vanishing $\frac{\partial\Gamma_c}{\partial w_c}$ whenever $V(X_c)=0$. Assuming $e^{Q_c}$ remains finite, this condition is equivalent to $w'(X_c)=0$, which is what we wanted to show.

The geometry near the fixed point is well-approximated by a constant dilaton vacuum, since the condition $V(X_c)=0$ is the defining relation of such vacua. The Ricci scalar according to \eqref{eq:chaos8} is positive if $V'(X_c)$ is positive. Thus, in order to get such a fixed point corresponding to dS$_2$ we need to ensure that $w(X_c)$ has not only an extremum, but rather a local maximum.

\begin{figure}
    \centering 
    \includegraphics[width=0.5\linewidth]{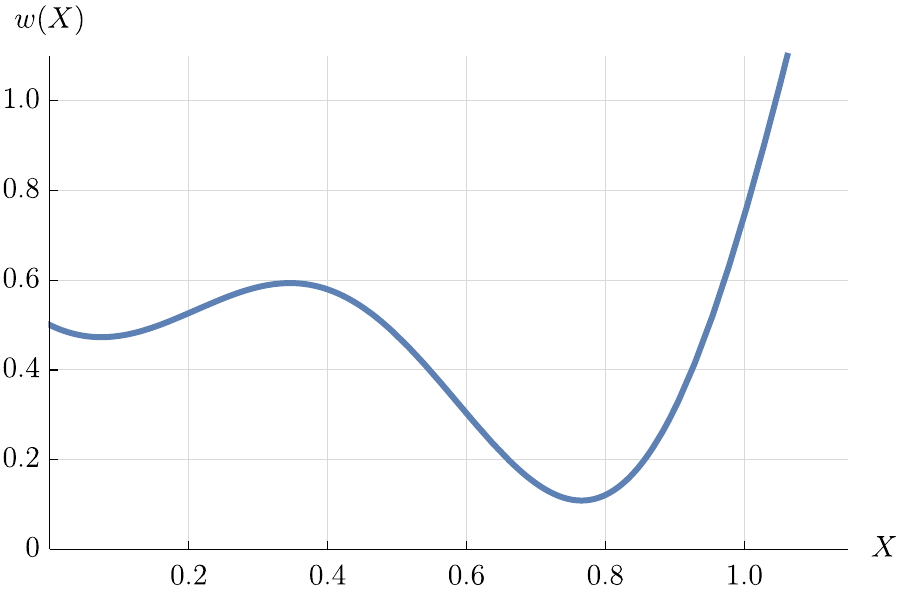}
    \caption{Random graph as example for $w(X)$ with asymptotic behavior $X^2$. Coming from infinity the first extremum on the right necessarily is a minimum and thus not a dS$_2$ fixed point. } 
    \label{fig:2}
\end{figure}

Now we see the incompatibility of the asymptotic AdS$_2$ behavior in the UV with the dS$_2$ fixed point in the IR: Since $w(X_c)$ asymptotically tends to $+\infty$ and is assumed to be sufficiently smooth (at least twice differentiable), the outermost extremum can only be a minimum. This means that even if $w(X_c)$ had a shape allowing for a local maximum and hence a dS$_2$ fixed point, it will never be reached by a smooth flow from the asymptotically AdS$_2$ region, since the first fixed point to be reached as we flow to the IR is another AdS$_2$ fixed point (or a flat space fixed point if the extremum degenerates into an inflection point). Figure \ref{fig:2} illustrates this for some smooth random function $w(X)$ that asymptotes to $X^2$.

\subsection{Centaur example and genericity of no-go}\label{sec:4.2}

The smooth ``centaur geometry'' example given in section 4 of \cite{Anninos:2017hhn}, $V(X)=\epsilon-\sqrt{X^2+\epsilon^2}$ with some finite $\epsilon$, is a special case that flows from AdS$_2$ at $X\to\infty$ to flat space at $X=0$. (To see this recall that according to \eqref{eq:chaos8} the Ricci scalar is given by $R=-2X/\sqrt{X^2+\epsilon^2}$.) The other asymptotic region, $X\to-\infty$, corresponds to a dS$_2$ region. But it cannot be reached by our flow, which necessarily terminates at the IR fixed point $X=0$. 

The conclusions of this section are rather generic and apply also to any model where $w(X)$ tends to $+\infty$, regardless of whether asymptotically AdS$_2$ is approached; the same line of arguments shows that the IR fixed point reached by smooth flows always is well-approximated by an AdS$_2$ or flat space fixed point and never by a dS$_2$ fixed point. Moreover, the no-go proof did not require a specific flow like \eqref{eq:chaos3}; instead, {\em any} flow associated with a change of the cutoff surface leads to the starting point \eqref{eq:chaos4} of our no-go result. 

\subsection{Evading the no-go?}\label{sec:4.3}

An immediate concern is whether our no-go result implies that AdS$_2$ in the UV cannot flow to positive curvature in the IR. As we show below, smooth flows do not reach a dS$_2$ \textit{fixed point} but they may still enter a region corresponding to positive curvature.

The only way to evade the no-go result we are able to find is if there is no IR fixed point at finite value of the dilaton, so that effectively the IR region is approached as $X\to-\infty$. This means that $w'$ must not vanish anywhere and hence $V$ is non-zero everywhere, which forbids in particular constant dilaton vacua. In that case an asymptotic dS$_2$ fixed point is possible. An example is provided by a small modification of the ``centaur'' potential above,
\eq{
V_{\textrm{\tiny{AdS}}}^{\textrm{\tiny{dS}}}(X) = -\sqrt{X^2+\eps^2}
}{eq:chaos11}
which flows to AdS$_2$ for $X\to\infty$ and to dS$_2$ for $X\to-\infty$. Then one can choose some $X_h <0$ with $|X_h| \gg \eps$ and construct a flow that moves past $X=0$ into the region $X<0$ where the Ricci scalar becomes positive.~\footnote{For this particular example one must choose $w_0$ in \eqref{omegadef} so that $w(X_c) > 0$ for $X_c>X_h$, in order to ensure reality of the on-shell action \eqref{OnShellAction1} along the flow.} Once $|X_c| \gg \eps$ the difference between the Ricci scalar and its dS$_2$ fixed-point value $R = 2$ is of order $(\eps/X_c)^2$.

However, there are two physical objections to a flow crossing the point $X=0$. First, in the models we study a vanishing dilaton effectively means infinite 2d Newton constant. And second, for $X_h<0$ the entropy $S = 2\pi X_h$ becomes negative. Both problems are resolved by including a topological term in the action \eqref{FullAction} of the form
\begin{align} \label{TopologicalFix}
    - \frac{1}{2}\,\int_{M} \nts\nts \extd^{2}x\,\sqrt{g}\,\bX\,R - \int_{\Sigma} \nts \extd^{1}x\,\sqrt{h}\, \bX\,K = -2 \pi \bX\,\chi 
\end{align}
for some positive constant $\bX$. This topological term for instance appears naturally in the dimensional reduction of nearly extremal black holes in higher dimensions, see e.g.~\cite{Almheiri:2014cka}. The solutions we consider have the topology of a disk ($\chi = 1)$, so the effect is a constant shift of the dilaton $X \to X + \bX$ in our previous results. (The functions $Q$, $w$, and $H$ are unaffected by the addition of this term in the action.) The condition that the 2d Newton constant remains finite is now $\bX + X > 0$, which also ensures that the entropy $S = 2\pi\,(X_h + \bX)$ remains non-negative for $|X_h| < \bX$. This modification allows $X$ to safely cross into the region $X_h < X<0$ as in the example above.

Therefore, our no-go result can be rephrased as the statement that any asymptotically AdS$_2$ geometry flows in the IR either to a fixed-point with non-positive constant curvature (AdS$_2$ or flat space), or it has no IR fixed point at all. In the latter case, the flow does not stop before reaching an endpoint at $X_h$ or moving into the strong coupling region. While we have not identified any loopholes, it is possible to construct flows that reach a region of positive scalar curvature. The example above modifies the potential $V(X)$ associated with the centaur geometry of \cite{Anninos:2017hhn} to eliminate a flat-space fixed point at $X=0$. Then the topological term in the action, which occurs naturally in holographic models and the dimensional reduction of solutions of string theory \cite{Almheiri:2014cka}, shifts the strong coupling region away from $X=0$ and allows the flow to move into the region $X<0$ where the curvature is positive and approximately constant. One can construct similar potentials $V(X)$ where the region $X<0$ corresponds to \emph{constant} positive curvature, but such flows still do not have an IR fixed point.

\section{Chaos}
\label{sec:Chaos}

The models described in section \ref{sec:Review} exhibit a variety of asymptotics as $X_c \to \infty$, yet they all satisfy a flow equation with the same general structure that emerges in dimensionally reduced $T\bar{T}$-deformations of AdS$_3$/CFT$_2$. In this section we consider saturation of the bound \eqref{eq:chaos1} on the Lyapunov exponent, another well-known feature of AdS/CFT, and ask whether the same behavior arises in a more general class of models. 

In section \ref{sec:5.1} we describe the calculation of the holographic Lyapunov exponent via the time delay in an outgoing signal due to an infalling shockwave. In section \ref{sec:5.2} we carry out the calculation for the full class of models described in section \ref{sec:Review} and find saturation of the bound \eqref{eq:chaos1}. This is followed by a brief discussion in \ref{sec:5.3}.

\subsection{Holographic Lyapunov exponent from time delay}\label{sec:5.1}

Consider a probe of the region near the horizon of a black hole of mass $M$ that consists of an infalling, massless particle of energy $\delta M \ll M$. This particle begins far from the horizon, moves inward along a null geodesic that crosses the surface $X_c$ at time $t_1$, and eventually falls into the black hole. An outgoing signal that crosses this shockwave at a point very near the horizon (or originates at that event) would be expected to arrive at $X_c$ at time $t_2$ in the original spacetime. However, the absorption of the shockwave increases the mass of the black hole by $\delta M$ and causes the horizon to grow. This leads to a delay in the outgoing signal, which arrives at $X_c$ at a later time $\tilde{t}_2$. We find that the delay $\tilde{t}_2 - t_2$ at $\mathcal{O}(\delta M/M)$ grows exponentially with the expected waiting time $t_2 - t_1$ and is characterized, for all the models considered in this paper, by the same maximal Lyapunov exponent that occurs in the asymptotically AdS$_2$ case.

The calculation of the delay in an outgoing signal due to an infalling shockwave closely follows the analyses in \cite{Engelsoy:2016xyb, Gross:2019ach}. The primary difference is our addition of a distant dilaton isosurface at $X_0 \gg X_c$ from which the infalling null matter originates. For models with AdS$_2$ asymptotics this surface can be removed to conformal infinity ($X_0 \to \infty$) with the infalling matter crossing $X_c$ after a finite amount of time. In models with different spacetime asymptotics, a finite $X_0$ eliminates the infinite waiting time required for matter to arrive from spatial infinity along an ingoing null geodesic.

%
%
\def \L {5} 
\definecolor{darkgreen}{HTML}{006622}
  \colorlet{triangle1}{darkgreen}
  \colorlet{triangle2}{orange}
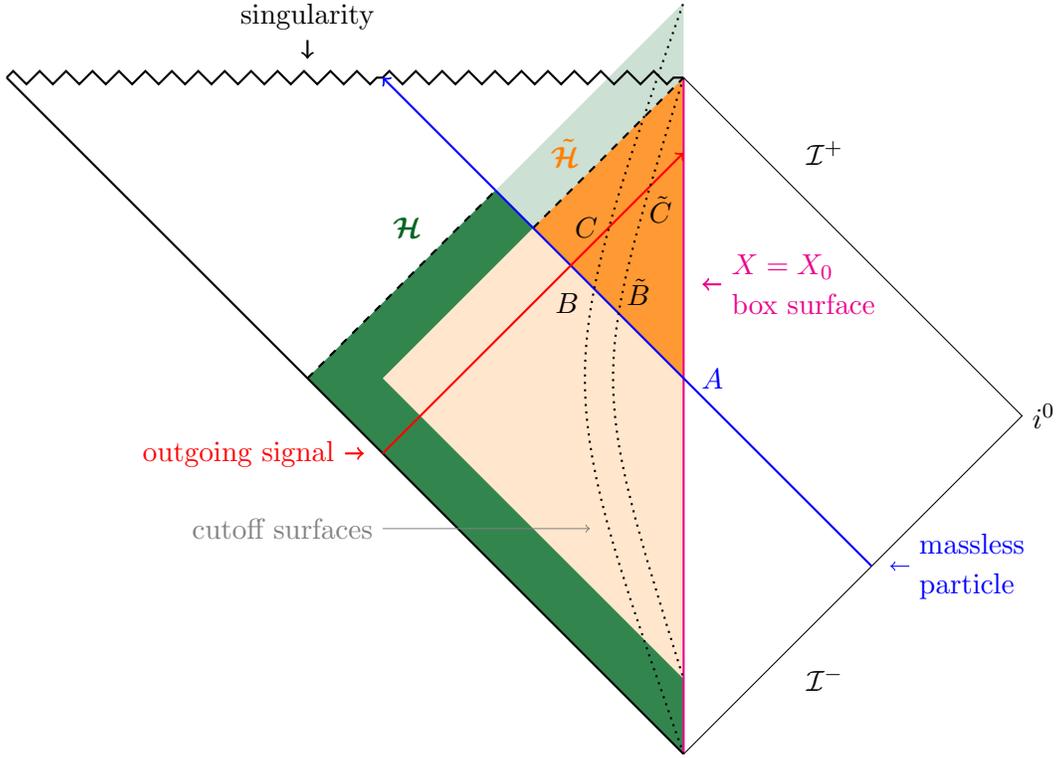
\begin{figure}
    \centering
\begin{tikzpicture}
%
  \draw[thick,black,zigzag] (-0.8*\L,0.8*\L) coordinate(stl) -- (0.2*\L,0.8*\L) coordinate (sin);
  \draw[thick,black,zigzag] (sin) -- (\L,0.8*\L) coordinate (end);
  \draw[thick,black,<-] (0.0*\L,0.85*\L) -- (0.0*\L,0.9*\L) node[above] {\color{black} singularity};
  \draw[thick,black] (\L,-\L) coordinate (sbr) -- (0,0) coordinate (bif) -- (stl);
  \draw[thick,black,dashed] (bif) -- (0.5*\L,0.5*\L) coordinate (mid);
  \draw[thick,black,dashed] (0.6*\L,0.4*\L) -- (end);
  \draw[black] (end) -- (1.9*\L,-0.1*\L) node[right] (io) {$i^0$} -- (sbr);
  \draw[black] (1.3*\L,0.6*\L) node[right]  (scrip) {$\mathcal{I}^+$}
               (1.3*\L,-0.8*\L) node[right] (scrip) {$\mathcal{I}^-$}
               (0.62*\L,0.6*\L) node[right] (scrip) {\color{triangle2}$\boldsymbol{\mathcal{\tilde{H}}}$}
               (0.2*\L,0.4*\L) node[right] (scrip) {\color{triangle1}$\boldsymbol{\mathcal{H}}$};          
 
  \colorlet{box}{magenta}
  \draw[thick,box]  (end) -- (\L,-\L) coordinate(bot);
  \draw[thick,box,<-] (1.05*\L,0.25*\L) -- (1.1*\L,0.25*\L) node[right,align=left] {$X=X_0$\\box surface};
  
  \draw[thick,blue] (1.5*\L,-\L/2) coordinate (scr) -- (\L,0) coordinate(ctr);
  \draw[thick,blue] (ctr) -- (\L/2,\L/2) coordinate(hor);
  \draw[thick,blue,->] (hor) -- (sin);
  \draw[thick,blue] (1.02*\L,0) node[right,align=left] {$A$};
  \draw[blue,<-] (1.55*\L,-0.5*\L) -- (1.6*\L,-0.5*\L) node[right,align=left] {massless\\ particle};
  
  \draw[thick,red,->] (0.2*\L,-0.2*\L) -- (\L,0.6*\L);
  \draw[thick,red,<-] (0.15*\L,-0.2*\L) -- (0.1*\L,-0.2*\L) node[left,align=left] {outgoing signal};
  \draw[thick,black] (0.75*\L,0.2*\L) node[left,align=left] {$B$};
  \draw[thick,black] (0.82*\L,0.23*\L) node[right,align=right] {$\tilde B$};
  \draw[thick,black] (0.8*\L,0.4*\L) node[left,align=left] {$C$};
  \draw[thick,black] (0.88*\L,0.45*\L) node[right,align=right] {$\tilde C$};
  
  \draw[thick,black,dotted] (sbr) .. controls (0.65*\L,0) .. (\L,\L) coordinate (str);
  \draw[thick,black,dotted] (end) .. controls (0.75*\L,0) .. (\L,-0.8*\L);
  \draw[very thin,gray,<-] (0.75*\L,-0.4*\L) -- (0.2*\L,-0.4*\L) node[left,align=left] {cutoff surfaces};
  
  \begin{pgfonlayer}{background}
   \fill[triangle1!20] (bif) -- (str) -- (sbr) -- cycle;
   \fill[triangle1!80] (bif) -- (mid) -- (\L,0) coordinate (ptA) -- cycle;
   \fill[triangle1!80] (bif) -- (ptA) -- (sbr) -- cycle;
   \fill[triangle1!80] (bif) -- (mid) -- (sbr) -- cycle;
   \fill[triangle2!20] (0.2*\L,0) -- (end) -- (\L,-0.8*\L) -- cycle;
   \fill[triangle2!80] (0.6*\L,0.4*\L) -- (end) -- (ptA) -- cycle;
  \end{pgfonlayer}
  
\end{tikzpicture}
    \caption{Shockwaves for asymptotically flat dilaton gravity in a box. Spacetime is put in a box at $X=X_0$. $A$ is the intersection of the box surface with the shockwave generated by a massless particle of energy $\delta M$. The horizon $\mathcal{H}$ grows to $\mathcal{\tilde H}$ after the shock. Extrapolations of spacetime regions beyond their regime of validity are weakly colored. $B$ ($\tilde B$) is the intersection of the shockwave with the initial (new) cutoff surface at time $t_1$ ($\tilde t_1$). $C$ ($\tilde C$) is the intersection of the outgoing signal with the initial (new) cutoff surface at time $t_2$ ($\tilde t_2$). See text for more explanations.}
    \label{fig:1}
\end{figure}

\subsection{Calculation for generic Maxwell-dilaton gravity}\label{sec:5.2}

As in section \ref{sec:dS} we set $f(X)=0$ to reduce clutter in the following calculation. The replacement $V \to V - q^2/(8f)$ (or equivalently, $w \to w + q^2\,H/4$) generalizes our result to models with $f(X) \neq 0$.

In the Schwarzschild-like coordinates used in \eqref{DGSolution} the line element takes the form
\begin{gather}
    \extd s^2 = -\xi(X)\,\extd t^2 + \frac{1}{\xi(X)}\,\extd r^2 ~.
\end{gather}
Using $\partial_r X = e^{-Q}$, null geodesics are given by 
\begin{gather}
    \extd t = \pm \frac{\extd X}{w(X)-2M} 
\end{gather}
where the $+$ and $-$ signs corresponds to outgoing and ingoing rays, respectively. The coordinate-time separation for two events $A$ and $B$ along a null geodesic is then
\begin{gather}
    t_B - t_A = \pm F(X_A,X_B) \qquad\qquad F(X_A,X_B) = \int\limits_{X_\ms{A}}^{X_\ms{B}}\frac{\extd X^\prime}{w(X')-2M} ~.
\end{gather}
This quantity is finite for two points with $X_h < X < \infty$, and exhibits a logarithmic divergence as one of the points approaches the horizon $X_h$. For models where $w$ grows faster than linear as $X \to \infty$, null rays reach or arrive from spatial infinity in finite time. But we consider a broader class of models, so we enclose the system in a ``box'' defined by the surface $X = X_0$ with $X_0 \gg X_c$. Figure \ref{fig:1} illustrates this construction for asymptotically flat dilaton gravity.

Now consider an ingoing null ray carrying energy $\delta M$ that starts at $X_0$ at $t=0$. It crosses $X_c$ at $t_1 = -F(X_0, X_c)$ and passes through a point $X_{\varepsilon} = (1+\varepsilon)X_h$ very near ($0 < \varepsilon \ll 1$) the horizon at time $t_\varepsilon = -F(X_0, X_\varepsilon) = t_1 - F(X_c,X_\varepsilon)
$. In the spacetime of a black hole with mass $M$, an outgoing ray passing through this event would reach $X_c$ at $t_2 = t_{\varepsilon} + F(X_{\varepsilon},X_c)$. So the expected waiting time between the ingoing and outgoing signals crossing $X_c$ is
\begin{gather}\label{NullRays}
    t_2 - t_1 = 2\,F(X_\varepsilon, X_c) ~.
\end{gather}
For $0 < \varepsilon \ll 1$ this waiting time is dominated by a $\log \varepsilon$ contribution from the region near the horizon.
\begin{gather}\label{t2minust1}
    t_2 - t_1 = \frac{2}{w_{h}{}'}\,\log \frac{1}{\varepsilon} + \ldots = \frac{\beta}{2\pi}\,\log \frac{1}{\varepsilon} + \ldots 
\end{gather}
The ellipsis indicates terms proportional to non-negative powers of $\varepsilon$. The proper time interval $T_2 - T_1$ measured by an observer at $X_c$ is related to this waiting time by a factor of $\sqrt{\xi_{c}}$, giving
\begin{gather}
    T_2 - T_1 = \frac{\beta_c}{2\pi}\,\log \frac{1}{\varepsilon} + \ldots \qquad \Rightarrow \qquad \frac{1}{\varepsilon} = \exp\left(\frac{2\pi}{\beta_c}\,(T_2 - T_1)\right) \times (\textrm{constant})
\end{gather}
The arrival of the shockwave increases the mass of the black hole to $\tilde{M} = M + \delta M$. Null rays in this spacetime satisfy
\begin{gather}\label{NullRaysTilde}
    \tilde{t}_{B} - \tilde{t}_{A} = \pm \tilde{F}(X_A,X_B)\qquad\qquad \tilde{F}(X_A,X_B) = \int\limits_{X_A}^{X_B} \frac{\extd X^\prime}{w(X')-2\tilde{M}} ~.
\end{gather}
We now repeat the calculation above, with the ingoing null geodesic passing through $X_{\varepsilon}$ at time $\tilde{t}_{\varepsilon}$ and an outgoing ray from that event crossing $X_c$ at time $\tilde{t}_{2}$. There is a shift in the location of the horizon given by $w(X_{h} + \delta X_{h}) = 2(M+\delta M)$, which to first order in $\delta M$ yields
\begin{gather}\label{deltaXh}
    \delta X_{h} = \frac{2\,\delta M}{w_{h}{}'} ~.
\end{gather}
As long as $\varepsilon > \delta X_{h} / X_{h}$ we may expand $\tilde{F}(X_A,X_B)$ in \eqref{NullRaysTilde} in powers of $\delta M$. The $\mathcal{O}(\delta M)$ shift in the time at which the outgoing null ray crosses $X_c$ is dominated by a $1/\varepsilon$ term when $0 < \varepsilon \ll 1$
\begin{gather}
    \tilde{t}_2 = t_2 + \frac{4\,\delta M}{(w_{h}{}')^2 X_h}\,\frac{1}{\varepsilon} + \ldots 
\end{gather}
where the ellipsis again denotes terms proportional to non-negative powers of $\varepsilon$. Thus, the proper crossing time $\tilde{T}_2$ as measured by an observer at $X_c$ in the spacetime with mass $\tilde{M} = M+\delta M$ differs from the expected value $T_2$ for an observer in the spacetime with mass $M$ by 
\begin{gather}\label{Delay}
    \tilde{T}_{2} - T_{2} = \frac{\beta_c}{2\pi}\,\frac{2\,\delta M}{w_{h}{}'\,X_{h}}\,\exp\left(\frac{2\pi}{\beta_c}\,(T_2 - T_1)\right) \times (\textrm{constant}) ~.
\end{gather}
Using \eqref{deltaXh} and the universal result $S = 2\pi X_{h}$ for the black hole entropy, this can be written as
\begin{equation}
    \tilde{T}_{2} - T_{2} = \frac{\beta_c}{2\pi}\,(\delta \log S)\,\exp\left(\frac{2\pi}{\beta_c}\,(T_2 - T_1)\right) \times (\textrm{constant}) ~.
    \label{eq:nolabel}
\end{equation}
The proper-time delay in the arrival of the signal grows exponentially with the expected waiting time and is characterized by the maximal Lyapunov exponent
\begin{gather}\label{Lyapunov}
    \lambda_L = 2\pi\, T_c 
\end{gather}
where $T_c = 1/\beta_{c}$ is the proper local temperature measured by an observer at $X_c$. In this calculation the proper times are defined with respect to the bulk metric at $X_c$. However, one may instead express the result in terms of proper quantities as measured with respect to the ``field theory'' metric of section \ref{sec:TbarT}, in which case the exponent is $\lambda_{L} = 2\pi\, T_\ts{FT}$.

\subsection{Chaos bound saturation}\label{sec:5.3}

The calculation of the time delay of an outgoing signal due to an infalling shockwave indicates that the bound \eqref{eq:chaos1} is saturated for all of the models described in section \ref{sec:Review}. This should not come as a surprise; the time delay \eqref{Delay} is due to the outgoing signal originating a bit closer to a horizon than it would have in the absence of a shockwave. This is a near-horizon effect that is insensitive to the asymptotics. As was pointed out in \cite{Gross:2019ach}, the time delay in the JT model is not affected by the presence of a finite cut-off at $X_c$. For the same reason, it is insensitive to both the ``box'' at $X_0 \gg X_c$ and the general behavior of $w(X)$ as $X \to \infty$. 

The benefit of the present calculation is that it establishes the result \eqref{Lyapunov} for a large group of models, in a manner that makes the insensitivity to the specific asymptotics manifest. On the other hand, there is no known holographic dual for most of the models we consider, which prevents us from matching the result \eqref{Lyapunov} to a boundary calculation.

\section{Discussion}
\label{sec:Discussion}

The general class of Maxwell-dilaton gravity models described in section \ref{sec:Review} shares certain universal features with the JT model. Specifically, they satisfy a flow equation with the same general structure as \eqref{eq:chaos3} and saturate the bound \eqref{eq:chaos1} on the Lyapunov exponent. For the JT model, these features are understood in the context of (dimensionally reduced) AdS$_3$/CFT$_2$ duality, and predictions on the dilaton gravity side can be matched to well-defined calculations in the dual quantum mechanics. But many of the Maxwell-dilaton gravity models we consider, which include the dimensional reduction of higher-dimensional theories with non-AdS asymptotics, do not have a known holographic dual. 

While our treatment of dilaton gravity was rather generic and included asymptotically AdS$_2$, Rindler or flat spacetimes we did make a couple of assumptions. The main motivation to relax some of them is the desire to further generalize the universality of the flow equation \eqref{UniversalFlowEquation} and chaos bound saturation \eqref{Lyapunov} (or to find counter examples). We review now our assumptions and reasons to drop them.

The first one concerns the boundary, which we assumed to be a dilaton isosurface. For applications to SYK/JT-like correspondences this assumption should be dropped. A consequence is an additional boundary term beyond the ones present in the action \eqref{FullAction}, namely a kinetic term for the dilaton \cite{Grumiller:2017qao}. This boundary term and the fact that the dilaton varies along the boundary will modify some of our discussion accordingly. We do not expect this generalization to change the flow equations, nor our main conclusions on chaos bound saturation. 

The second one is asymptotic positivity of the Weyl invariant dilaton function $w$ \eqref{omegadef}, which excludes in particular asymptotic dS$_2$ behavior.\footnote{%
One of the key results of the study of two-dimensional cosmological horizons with scalar matter in \cite{Anninos:2018svg} is the absence of a Lyapunov exponent. Instead, their out-of-time-ordered correlation functions oscillate as a consequence of a crucial sign change (their kinetic term in the Schwarzian action has the `wrong' sign). Their results about cosmological horizons do not contradict our results about black hole horizons, but they showcase that one has to be careful in extrapolating black hole results and insights to cosmological ones.
} For toy models of cosmology it could be useful to avoid this assumption. As we showed in section \ref{sec:3.2} the flow equation \eqref{FinalFlowEquation} follows from the quasi-local form of the first law of black hole thermodynamics. Theories that admit cosmological horizons also satisfy a first law, so it is not unreasonable to expect that it can also be recast as a flow equation. However, without relaxing our condition on $w$ we cannot speculate on the structure and interpretation of such a flow equation, if it exists, and we therefore leave this point open to future discussions.

The third assumption is asymptotic dominance of $w$ over the coupling function $H$ \eqref{Hdef}. Relaxing this requirement would allow, among other things, to discuss models where the cosmological constant is state-dependent \cite{Grumiller:2014oha}. Another generalization in the same spirit, called `asymptotic mass domination' was discussed in \cite{Bagchi:2014ava}. In that case neither $w$ nor $H$ but rather the mass $M$ in the Killing norm \eqref{MetricSolution} dominates the asymptotic behavior. Both of these cases require a modification of the boundary action \eqref{FullAction}. We have not checked how such generalizations affect the flow equations. It could be rewarding to do so.

Even if we drop none of the assumptions above there are numerous generalizations by including extra fields, such as matter fields, extra (non-)abelian gauge fields or higher spin fields. The impact of matter fields on the interpretation of $T\bar T$-deformations \cite{Guica:2019nzm} are indicative that the inclusion of matter fields can lead to qualitative changes, in particular concerning the interpretation of the dual field theory. It could be worthwhile to investigate which of the generalizations has an impact on the flow equations, the chaos bound saturation and the (existence of the) dual field theory. 

Finally, we stress that our no-go result does not imply the absence of potentially interesting IR dynamics in the dual field theory. While there is no way to connect an AdS$_2$ fixed point in the UV with a dS$_2$ fixed point in the IR, we have confirmed that centaur-inspired flows like the one associated with \eqref{eq:chaos11} exist, where asymptotically AdS$_2$ flows into a strong coupling region of (approximately) positive curavture. If there was a dual field theory it would exhibit QCD-like behavior in the sense that from the dual field theory perspective we have a (nearly) CFT fixed point in the UV that flows to a strong coupling region in the IR.

\section*{Acknowledgments}

DG was supported by the Austrian Science Fund (FWF), projects P~30822 and P~32581.
RM thanks Loyola University Chicago for support via a Summer Research Stipend.

\bibliographystyle{fullsort}
\bibliography{review}

\end{document}